\documentclass[aps,twocolumn,prb,tightenlines,floatfix,showpacs,nofootinbib]{revtex4}
\usepackage[dvips]{graphicx}
\usepackage[english]{babel}
\usepackage{amsmath}
\usepackage{amssymb}
\usepackage{bm}
\usepackage{color}
\usepackage{subfigure}

\newcommand{\be}{\begin{eqnarray}}
\newcommand{\ee}{\end{eqnarray}}
\newcommand{\vk}{\mathbf{k}}
\newcommand{\vn}{\mathbf{n}}
\newcommand{\p}{\partial}

\begin{document}

\title{The $Z_2$ Classification of Dimensional Reduced Hopf Insulators}
\author{Chang-Yan Wang}
\affiliation{College of Physical Science and Technology,
Sichuan University, Chengdu, Sichuan 610064, China}
\author{Yan He}
\affiliation{College of Physical Science and Technology,
Sichuan University, Chengdu, Sichuan 610064, China}

\begin{abstract}
The Hopf insulators are characterized by a topological invariant called Hopf index which classifies maps from three-sphere to two-sphere, instead of a Chern number or a Chern parity. In contrast to topological insulator, the Hopf insulator is not protected by any kind of symmetry. By dimensional reduction, we argue that there exists a new type of $\mathbb{Z}_2$ index for 2D Hamiltonian with vanishing Chern number. Specific model Hamiltonian with this nontrivial $\mathbb{Z}_2$ index is constructed. We also numerically calculate the topological protected edge modes of this dimensional reduced Hopf insulator and show that they are consistent with the $\mathbb{Z}_2$ classification.

\end{abstract}

\maketitle

\section{introduction}

In recent years, topological insulator and superconductor have attracted a lot of attentions both theoretically and experimentally\cite{Kane rev,Zhang rev}. The theoretical study started from the $\mathbb{Z}_2$ classification of 2D time reversal invariant models proposed by Kane\cite{Kane}, and then generalized to 3D topological insulator and superconductor\cite{Fu,Moore}. The topological property can be revealed by a nonzero topological index when the model is defined on a closed manifold. Alternatively, it is also reflected by the existence of edge modes when the model is defined on a manifold with boundaries.

At first, these results looks quite scattered. A few years ago, by making use of K theory, Kitaev pointed out that the topological states can be organized into a periodic table which classifies the tight-banding Hamiltonians in any dimensions\cite{Kitaev}. There are ten types of Hamiltonians defined according to the presence or absence of certain discrete symmetries such as time reversal, particle-hole and chiral symmetry. The periodic table shows that there are 5 topological nontrivial types in each dimension. This classification actually reflected the Bott periodicity\cite{Stone} of the homotopy groups of the classical Lie groups such as $U(N)$, $SO(N)$ and $Sp(N)$. It is well known that the Bott periodicity only happens to stable homotopy groups of Lie groups with large enough ranks. Therefore, for low dimensional models with a few number of bands, there may be exceptions to the above mentioned topological periodic table.

An interesting exception is the 3D Hopf insulator\cite{Wen}. The space of a general band insulator with $m$ filled bands and $n$ empty bands is topologically equivalent to the Grassmannian manifold $\mathbb{G}_{m,m+n}$. In 3D, these band insulators can be classified by  maps $T^3 \to \mathbb{G}_{m,m+n}$. But the homotopy group $\pi_3(\mathbb{G}_{m,m+n})=\{0\}$ expect for $m=n=1$, which mean there are no nontrivial topological states in these cases. However, for $m=n=1$, $\mathbb{G}_{m,m+n}$ is topologically equivalent to $S^2$ and the Hopf map from $S^3$ to $S^2$ gives a nontrivial homotopy group $\pi_3(S^2)=\mathbb{Z}$. This indicates that the existence of nontrivial topological states in two band 3D insulators and specific tight-banding Hamiltonians with arbitrary Hopf index has indeed been constructed\cite{Deng}.

In this paper, we propose a new type of $\mathbb{Z}_2$ classification for 2D two-band models with vanishing Chern number. We following closely the method of dimensional reduction\cite{Zhang} which has been used to establish the $\mathbb{Z}_2$ classification of 3D topological insulator staring from a 4D Hamiltonian with nonzero second Chern number. We noticed that the dimensional reduction actually gives rise to a higher order homotopy group. Similarly, the dimensional reduction of Hopf insulator also corresponds to the homotopy group $\pi_4(S^2)=\mathbb{Z}_2$. We will call this topological nontrivial state as dimensional reduced Hopf insulator. This state clearly sits outside the periodic table of topological insulators and it is not protected by any discrete symmetries.

The rest of this paper is organized as follows. In section \ref{Hopf}, we review the concept of the Hopf index and the construction of the Hopf insulator. The $\mathbb{Z}_2$ classification is established by the dimensional reduction in section \ref{reduce}. In section \ref{num}, we present the numerical results of edge modes. We conclude in section \ref{conclude}.

\section{Hopf index and Hopf insulator}
\label{Hopf}

For a general two band insulators, one can always expand its Hamiltonian in terms of Pauli matrices. In momentum space, the Hamiltonian can be written as
\begin{equation}
H(\mathbf{k})=\sum_{a=1}^3 d_a(\mathbf{k})\sigma ^a +\epsilon(\mathbf{k})\mathbb{I}\label{Hk}
\end{equation}
where $\mathbb{I}$ is the $2 \times 2$ identity matrix, $\sigma ^a $ are the three Pauli matrices and $\vk=(k_x,k_y,k_z)$. The $\epsilon(\mathbf{k}) \mathbb{I}$ term can be dropped without changing the topological properties, since it only shifts the overall energy levels. After omitting this term,the energy spectrum can be easily obtained as
\be
E_\pm=\pm\sqrt{\sum_a d_a ^2(\vk)}
\ee
In order to keep the energy gap open, it is required that $|E_\pm|>0$ or the $\mathbf{d}$ cannot vanish for any $\vk$. We notice that the Pauli matrices are just the basis of $\mathfrak{so}(3)$ (the Lie algebra of SO(3)), and this Hamiltonian all has a U(1) (equivalent to SO(2)) redundant freedom. Hence all possible band Hamiltonians form the space $SO(3)/SO(2)\equiv S^2$\cite{Naka}, which means equation (\ref{Hk}) gives a map from the momentum space to $S^2$. To make the picture more clear, we can introduce
\be
\vn(\vk)=\frac{(d_1(\mathbf{k}), d_2(\mathbf{k}), d_3(\mathbf{k}))}{\sqrt{d_1(\mathbf{k})^2+d_2(\mathbf{k})^2+d_3(\mathbf{k})^2}}.
\ee
Then the vector $\vn$ forms a 2D sphere in $\mathbb{R}^3$. Now it is clear that the classification of two band insulators in 2D is just the classification of the map $T^2\to S^2$, which is equivalent to the homotopy group $\pi_2(S^2)$ classifying the map $S^2\to S^2$, because the nontrivial cycles on the torus do not affect the homotopy classes of the map. It is well known that the winding number of this map is the same as the first Chern number which is the only topological invariant that classifies 2D Hamiltonian without any symmetry\cite{Simons}. This result has been generalized to higher dimensions. Since the first Chern number can only be defined as a 2D surface integral, the only topological invariants in higher dimension are the Chern numbers defined on all possible independent 2D subspaces. But there is an important exception in 3D. Actually, in 3D, the Hopf index provide a even finer topological classification than the Chern number.

In the following, we will describe a general definition of the Hopf index. There is a subtle difference between the map $T^3\to S^2$ and the map $S^3\to S^2$, but we will postpone the discussion of this point later and focus on the homotopy $\pi_3(S^2)$ for now. For the map $\vn$ defined before, we can define winding number density on $S^2$ as
\be
F_{\mu\nu}=\vn\cdot(\p_{\mu}\vn\times\p_{\nu}\vn)\label{F1}
\ee
Here $\mu,\nu=x,y,z$ and $\p_{\mu}=\frac{\p}{\p k_{\mu}}$. For two band models, the winding number is equivalent to the Chern number, thus the above defined winding number density is the same as the Berry curvature. In general, this winding number density which is also a 2-form (anti-symmetric 2nd order tensor) cannot be globally written as a derivative of a vector field. Because if this happens, the surface integral of the 2-form on the $S^2$ will be zero by Stokes theorem, which leads to a contradiction. But the map $T^3\to S^2$ provide a pullback of the 2-from back to $T^3$. On the $T^3$, we can indeed write the 2-form as a derivative of certain vector field globally
\be
F_{\mu\nu}=\partial_\mu A_\nu-\partial_\nu A_\mu\label{F2}
\ee
Here $A_\mu$ behaves like a $U(1)$ gauge field, which is not uniquely defined. Different $A_{\mu}$ related by $U(1)$ gauge transformation will give rise to the same 2-form $F_{\mu\nu}$. With the above defined $A_{\mu}$ and $F_{\mu\nu}$, the Hopf index can be expressed as an integral of Chern-Simons forms\cite{Wilczek}
\be
\mathcal{H}=\frac{1}{32\pi^2}\int_{BZ}\epsilon_{\mu\nu\rho}A_{\mu}F_{\nu\rho} d^3k
\label{eq:HI}
\ee
Note that the expression of Hopf index depends on $A_{\mu}$ , but Hopf index is gauge invariant. Suppose we make an infinitesimal gauge transformation $A_{\mu}\to A_{\mu}-\p_{\mu}f $, we find the change of the Hopf index is
\be
\delta\mathcal{H}&=&\frac{1}{32\pi^2}\int_{BZ}\epsilon_{\mu\nu\rho}(\p_{\mu}f)F_{\nu\rho} d^3k\nonumber\\
&=&-\frac{1}{32\pi^2}\int_{BZ}f\epsilon_{\mu\nu\rho}\p_{\mu}F_{\nu\rho} d^3k=0
\ee
In the last step, we have used Bianchi identity.

Now we consider the case that the momentum $\vk$ is taken value form a 3D torus $T^3$. The maps from $T^3$ to $S^2$ is more complicated\cite{Fox} than the the maps from $S^3$ to $S^2$, because there are three independent 2D torus inside $T^3$ and each of them may has nonzero Chern numbers. For example, $C_x$ is the Chern number when taking $k_x$ to be constant.  For this complicated case, we will only quote the mathematical results. It is pointed out by Pontryagin\cite{Pont} that if $C_{\mu}\neq0$, then Hopf index is no longer integer-valued and take values in the finite group $\mathbb{Z}_{2\cdot GCD(C_x, C_y, C_z)}$, where GCD denotes the greatest common divisor. If the Chern numbers $C_\mu$ vanishes in all three directions, then the Hopf index will be integer-valued as in the $S^3$ case. In this paper, we only consider the later case with vanishing Chern numbers.

Next, we will follow Ref.\cite{Deng} to construct a class of 3D two band Hamiltonians with arbitrary integer number of Hopf index $\mathcal{H}$ and zero Chern number $C_\mu$ on the three sub-2D tori. First we define
\be
&&u_1(\mathbf{k})=(\sin k_x+i\sin k_y)^p,\nonumber\\
&&u_2(\mathbf{k})=[\sin k_z +i(\cos k_x + \cos k_y + \cos k_z +h)]^q\nonumber
\ee
then the 3D two band Hamiltonian is given by
\be
H=\sum_{i=1}^3d_i\sigma^i,\quad d_i=\sum_{a,b=1}^2u^*_a \sigma^i u_b\label{Model}
\ee
here $^*$ means complex conjugate, $h$ is a constant parameter and $p,q$ are two coprime integers. This Hamiltonian actually defines a map from $T^3$ to $S^2$. To see this more clearly, we can normalize $u_a$ to introduce
\be
z_a=\frac{u_a}{\sqrt{|u_1|^2+|u_2|^2}},\quad(a=1,2)
\ee
Then one can easily see that $|z_1|^2+|z_2|^2=1$ which describe a unit 3D sphere in $\mathbb{R}^4$. Thus $z_a(\vk)$ gives a map from $T^3$ to $S^3$. We can also define the normalized $d_i$ by $d_i=\sum_{a,b=1}^2z^*_a \sigma^i z_b$ or more explicitly by
\be
d_1=\mbox{Re}(2z_1z_2^*),\,\, d_2=\mbox{Im}(2z_1z_2^*),\quad
d_3=|z_1|^2-|z_2|^2\nonumber
\ee
It is easy to verify that $\sum_{i=1}^3d_i^2=1$ which describe a unit 2D sphere in $\mathbb{R}^3$. Thus $d_i$ give a map from $S^3$ to $S^2$. This is the Hopf map discovered long time ago by mathematicians\cite{Whitehead}. The composition of the above two maps gives the desired map from $T^3$ to $S^2$ with nonzero Hopf index.

For this specific map, we have explicit formula for both the gauge field and field strength\cite{Fradkin}
\be
&&A_{\mu}=-i\sum_a\Big[z_a^*(\p_{\mu}z_a)-(\p_{\mu}z_a^*)z_a\Big]\nonumber\\
&&F_{\mu\nu}=-2i\sum_a\Big[(\p_{\mu}z_a^*)(\p_{\nu}z_a)-(\p_{\nu}z_a^*)(\p_{\mu}z_a)\Big]\nonumber
\ee
And Hopf index is still given by Eq.(\ref{eq:HI}). For the simplest case when $p=q=1$, we find
\be
\mathcal{H}&=&\frac{1}{32\pi^2}\int_{BZ}\frac{16(s_2+h s_3)}{(3+h^2+2s_2+2hs_1)^2}d^3k\nonumber\\
s_1&=&\cos k_x+\cos k_y+\cos k_z\nonumber\\
s_2&=&\cos k_x\cos k_y+\cos k_y\cos k_z+\cos k_z\cos k_x\nonumber\\
s_3&=&\cos k_x\cos k_y\cos k_z\nonumber
\ee

For $h=-3/2$, we numerically evaluate this integral and find that in this case $\mathcal{H}=1$. For more general $p,q$, the integrand is very complicated. One can discretize the integral and find that the Hopf index is $\pm pq$ where the sign depends on the orientation of $S^3$. This result agrees with the one from mathematical literature\cite{Whitehead}. In the above model with fixed $k_x$, one can verify that the Berry curvature satisfies $F_x(k_y,k_z)=-F_x(-k_y,-k_z)$, thus the Chern number in this direction is zero\cite{Deng}
$C_x=\int dk_ydk_z F_x(k_y,k_z)$. Similarly, we also have $C_y=C_z=0$, therefore for the model of Eq.(\ref{Model}), the Hopf index takes values in $\mathbb{Z}$ not in any finite group.

\section{dimensional reduction of Hopf insulator}
\label{reduce}

The method of dimensional reduction has been used to classify time reversal invariant topological insulators\cite{Zhang}. The time reversal invariant Hamiltonians belongs to the $Sp(N)$ group. In real space, the eigenstates are double degenerate to form Kramers pairs. There is a $Sp(1)$ transformation factor between any
two Kramers pairs. In 4D, it is known long time ago that time reversal invariant Hamiltonians is classified by second Chern number which reflect the the homotopy group $\pi_3(sp(1))=\mathbb{Z}$. By dimensional reduction, one can deduce the $\mathbb{Z}_2$ classification of 3D topological insulators, which reflect the homotopy group $\pi_4(sp(1))=\mathbb{Z}_2$. We observe that the procedure of dimension reduction corresponds to a shifting of the order of the homotopy group. Similarly, the integer index of 3D Hopf insulator reflects the homotopy group $\pi_3(S^2)=\mathbb{Z}$, and we can expect that there should be a $Z_2$ index if we dimensional reduce the 3D Hopf insulator, because $\pi_4(S^2)=Z_2$.

By dimensional reduction, we will establish the $\mathbb{Z}_2$ topological classification of 2D two band Hamiltonians with vanishing Chern number. We consider two different generic 2D two band Hamiltonians $H_1(\vk)$ and $H_2(\vk)$ which breaks time reversal symmetry but has zero first Chern number. Here $\vk=(k_x,k_y)$. We can always define a continuous path $p(\vk,\theta),\ \pi\in[0,\pi]$ interpolating these two Hamiltonians such that $p(\vk,0)=H_1(\vk)$ and $p(\vk,\pi)=H_2(\vk)$. The interpolation $p(\vk,\theta)$ is certainly not unique. First, we will show that for each interpolation path, one can define a Hopf index which associates with this path. Second, we will show that the difference between the Hopf indices of two different interpolation is always an even integer. These two steps completes the $\mathbb{Z}_2$ classification of the interpolations.

For interpolation $p(\vk,\theta)$, we define another continuous interpolation by
$$
q(\vk,\theta)=U^{\dagger}(\theta)p(\vk,\theta)U(\theta)
$$
which can be thought as a gauge transformation of original path $p(\vk,\theta)$. Here $U(\theta)$ is $\theta$ dependent unitary matrix which also satisfies $U(0)=U(\pi)$. Thus $U(\theta)$ is actually a loop in the gauge transformation space. Now we can put these two pathes together to form a closed loop in the Hamiltonian space
\be
P(\vk,\theta)=\begin{cases}
p(\vk,\theta)& \theta\in[0,\pi]\\
q(\vk,2\pi-\theta)& \theta\in[\pi,2\pi]
\end{cases}\label{loop}
\ee
Since $P(\vk,0)=P(\vk,2\pi)$, we can treat parameter $\theta$ as another momentum component. Thus $P(\vk,\theta)$ actually gives a map from $T^3$ to $S^2$. We can again expand $P(\vk,\theta)$ as $P(\vk,\theta)=\sum_a d_a(\vk,\theta)\sigma^a$, then the Berry curvature and Berry phase $F_{\mu\nu}$, $A_{\mu}$ are again calculated by Eq. (\ref{F1}), (\ref{F2}). The Hopf index associated with path $P(\vk,\theta)$ is
\be
\mathcal{H}[P(\vk,\theta)]=\frac{1}{32\pi^2}\int_0^{2\pi}d\theta\int_{BZ}d^2k\epsilon_{\mu\nu\rho}A_{\mu}F_{\nu\rho}
\ee
Since the choice of $U(\theta)$ is not unique, it may looks like that $\mathcal{H}[P(\vk,\theta)]$ is not uniquely defined. But actually, for different choices of $U(\theta)$, the corresponding $q(\vk,\theta)$ are related by unitary transformation. Therefore the corresponding Berry curvature are the same. Although the Berry phase may depend on gauge choice, it is known that Hopf index is gauge invariant. Therefore, one can see that $\mathcal{H}[P(\vk,\theta)]$ associated with path $P(\vk,\theta)$ is unique.

\begin{figure} \centering
\subfigure[] { \label{inter:a}
\includegraphics[width=0.45\columnwidth]{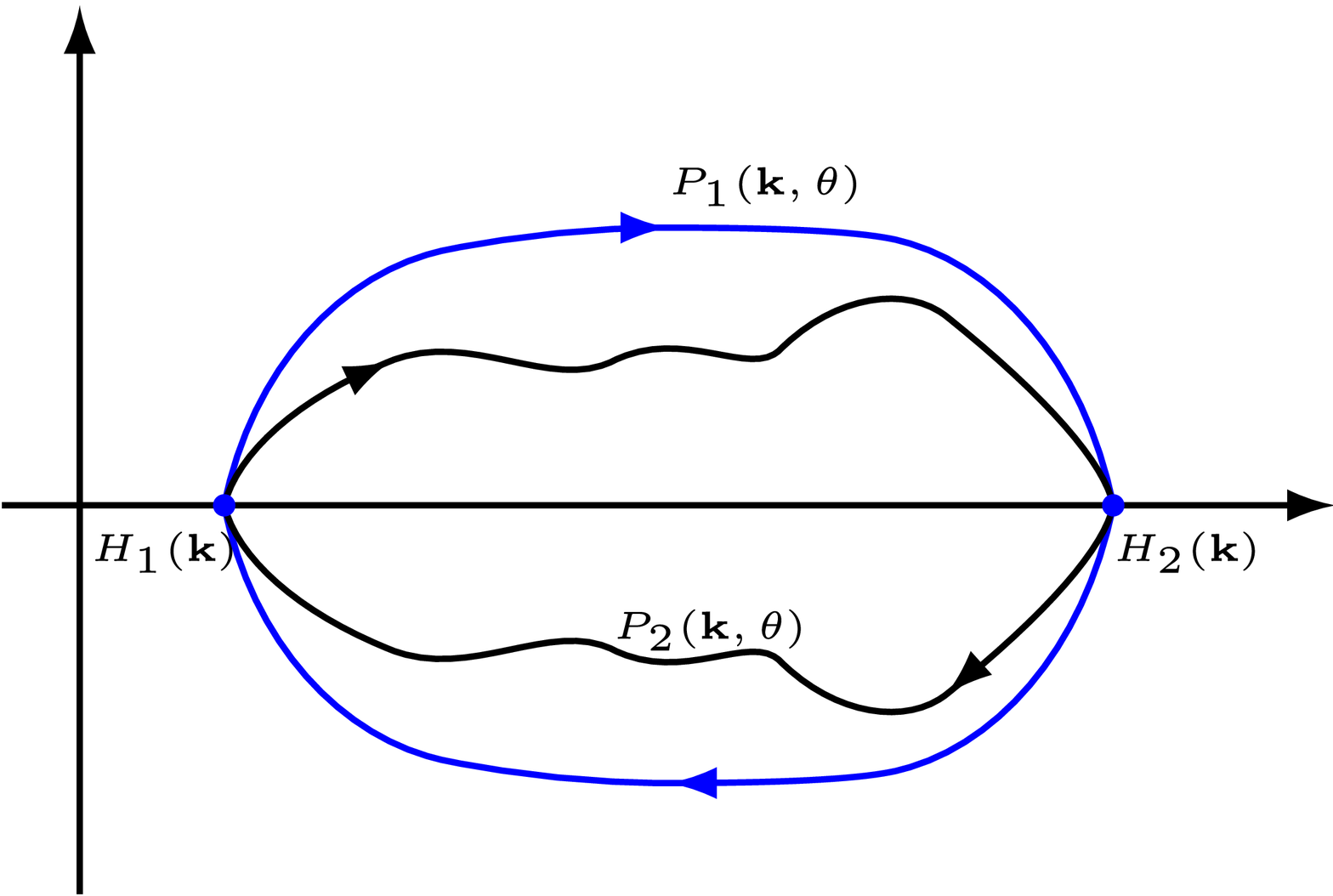}
}
\subfigure[] { \label{inter:b}
\includegraphics[width=0.45\columnwidth]{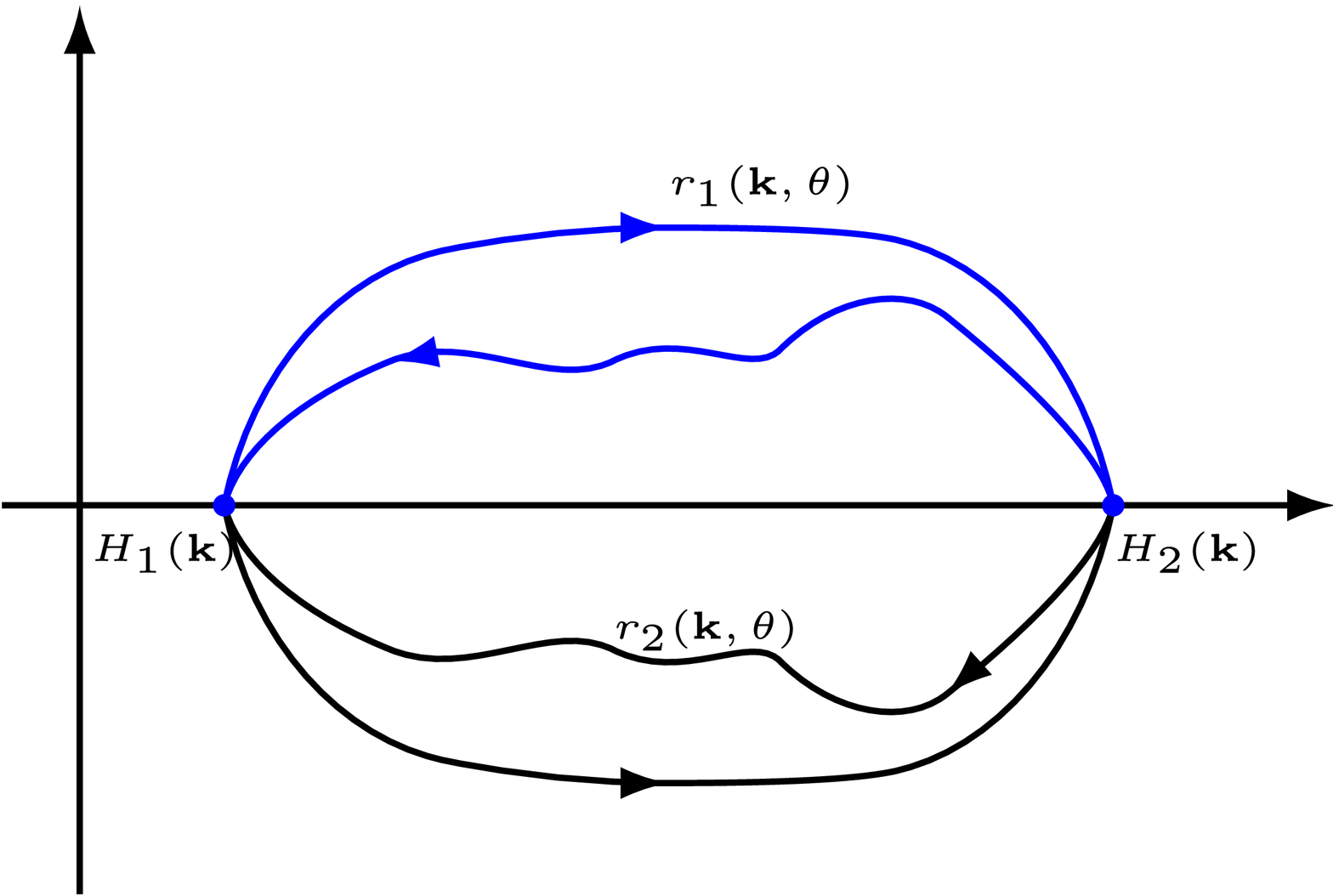}
}
\caption{Loops between $H_1(\vk)$ and $H_2(\vk)$. (a) The blue loop stands for $P_1(\vk,\theta)$, the black loop stands for $P_2(\vk,\theta)$. (b) The blue loop stands for $r_1(\vk,\theta)$, the black loop stands for $r_2(\vk,\theta)$.}
\label{inter}
\end{figure}

Now we consider two different interpolation $p_1(\vk,\theta)$ and $p_2(\vk,\theta)$ between the two Hamiltonians $H_{1,2}(\vk)$. For each of them, we define
$$
q_a(\vk,\theta)=U^{\dagger}_a(\theta)p_a(\vk,\theta)U_a(\theta),\quad (a=1,2)
$$
Then the path $p_{1,2}$ can be extended to become loops $P_{1,2}(\vk,\theta)$ as we discussed above, as shown in Fig. \ref{inter:a}. We can also form two different loops as
\be
r_1(\vk,\theta)=\begin{cases}
p_1(\vk,\theta)& \theta\in[0,\pi]\\
p_2(\vk,2\pi-\theta)& \theta\in[\pi,2\pi]
\end{cases}\\
r_2(\vk,\theta)=\begin{cases}
q_1(\vk,\theta)& \theta\in[0,\pi]\\
q_2(\vk,2\pi-\theta)& \theta\in[\pi,2\pi]
\end{cases}
\ee
as shown in Fig. \ref{inter:b}.
It is straightforward to see that the Hopf index of these 4 loops satisfy
\be
\mathcal{H}[P_1(\vk,\theta)]-\mathcal{H}[P_2(\vk,\theta)]=\mathcal{H}[r_1(\vk,\theta)]+\mathcal{H}[r_2(\vk,\theta)]\label{HH}
\ee
We see that $r_2(\vk,\theta)=U(\theta)^{\dagger}r_1(\vk,\theta)U(\theta)$ by definition. If we expand
$$
r_{i}=\sum_{a=1}^3 d^{i}_a\sigma^a \quad (i=1,2)
$$
then the two normalized 3-vector $\vn_i=\mathbf{d}^i/|\mathbf{d}^i|$ for $i=1,2$ are related by orthogonal transformation. It is well known that the area 2-form $\vn\cdot(\p_{\mu}\vn\times\p_{\nu}\vn)$ is invariant under the orthogonal transformation. Therefore, $F_{\mu\nu}$ are the same for both $r_1(\vk,\theta)$ and $r_2(\vk,\theta)$. Corresponding to the same $F_{\mu\nu}$, $A_{\mu}$ can take different forms depending on the gauge choice. But Hopf index is gauge invariant, thus we find that
$$
\mathcal{H}[r_1(\vk,\theta)]=\mathcal{H}[r_2(\vk,\theta)]
$$

Combining the above relation with Eq.(\ref{HH}), we arrive at the promised result
\be
\mathcal{H}[P_1(\vk,\theta)]-\mathcal{H}[P_2(\vk,\theta)]=0\mod2
\ee
This means that all possible interpolations between two given Hamiltonians can be divided into two classes.
If $\mathcal{H}[P(\vk,\theta)]$ is even, then the two Hamiltonians is topologically the same. On the other hand, if
$\mathcal{H}[P(\vk,\theta)]$ is odd, then the two Hamiltonians are topologically different and cannot be continuously deformed to each other. Therefore If we take the vacuum or its equivalent as a reference Hamiltonian, then the two Band 2D Hamiltonians with vanishing Chern number can be divided into two class. One can define $(-1)^{\mathcal{H}[P]}$ as a new type of $\mathbb{Z}_2$ index which provides a finer classification than Chern number. The topological nontrivial class will be called dimensional reduced Hopf insulator. This new type of topological insulator is quite different from quantum spin Hall state in 2D, although they are both classified by $\mathbb{Z}_2$ index. The quantum spin Hall state is protected by time reversal invariance and requires minimal 4 bands (2 orbital and 2 spins), while Hopf insulator is not protected by any symmetry and is realized in a 2 band model.

The above discussion is quite abstract. We will apply this discussion to the 3D Hopf Hamiltonian defined in Eq.(\ref{Model}). We define the following 2D Hamiltonians $H_1=H(\vk, k_z=0)$ and $H_2=H(\vk, k_z=\pi)$ where $\vk$ is 2-vector of $xy$ components and $H$ is the 3D Hamiltonian defined in Eq.(\ref{Model}). We can regard the $H(k_x, k_y, k_z)$ as the interpolation between $H_1$ and $H_2$. It is easy to see that the eigenvalues of $H(k_x, k_y, k_z)$ is
\be
E=\pm\sqrt{|u_1|^2+|u_2|^2}
\ee
Thus $H(\vk, k_z)$ and $H(\vk, -k_z)$ have the same eigenvalues and are related by unitary transformations. We define the following closed loop
\be
r(\vk,\theta)=\begin{cases}
H(k_x,k_y,\theta)& \theta\in[0,\pi]\\
H(k_x,k_y,\theta-2\pi)& \theta\in[\pi,2\pi]
\end{cases}
\ee
which satisfies the same conditions as in Eq.(\ref{loop}). The associated Hopf index $\mathcal{H}[r(\vk,\theta)]$ is just the Hopf index we have calculated for the 3D model Eq.(\ref{Model}). For the parameters $p=q=1$ and $h=-3/2$, we have $\mathcal{H}[r]=1$. In section \ref{num}, we find that $H_2$ is topologically trivial, thus $H_1$ is the desired example of dimensional reduced Hopf insulator.

\begin{figure} \centering
\subfigure[] { \label{fig:a}
\includegraphics[width=0.45\columnwidth]{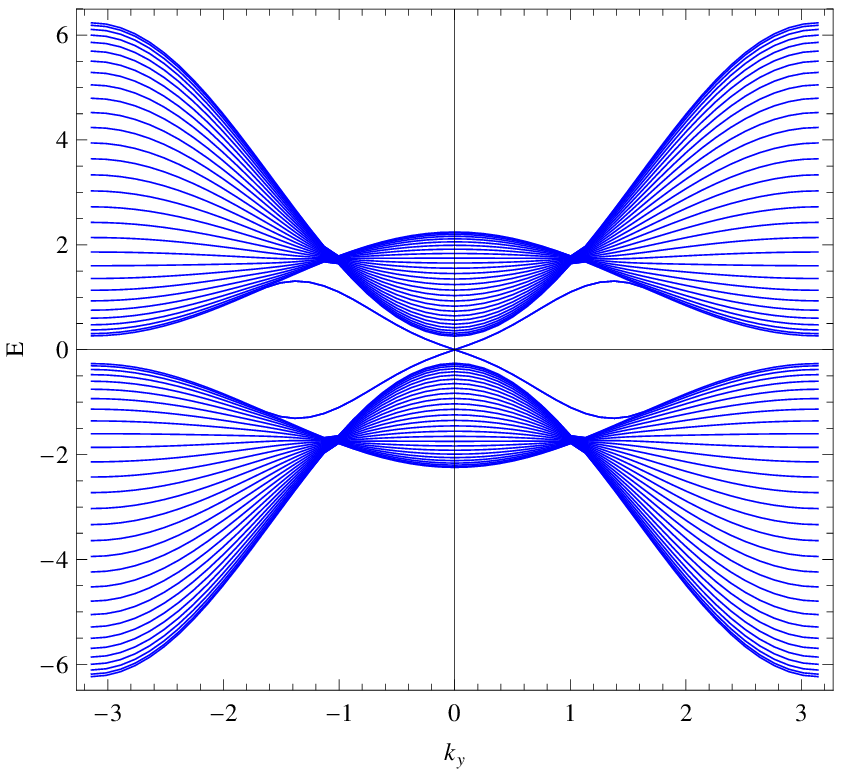}
}
\subfigure[] { \label{fig:b}
\includegraphics[width=0.45\columnwidth]{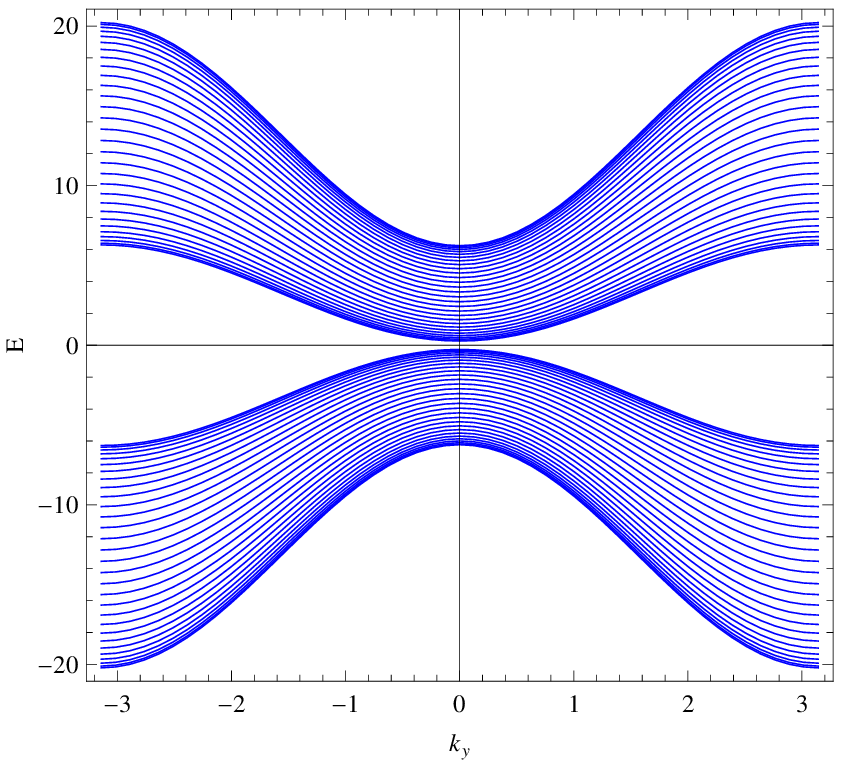}
}
\subfigure[] { \label{fig:c}
\includegraphics[width=0.45\columnwidth]{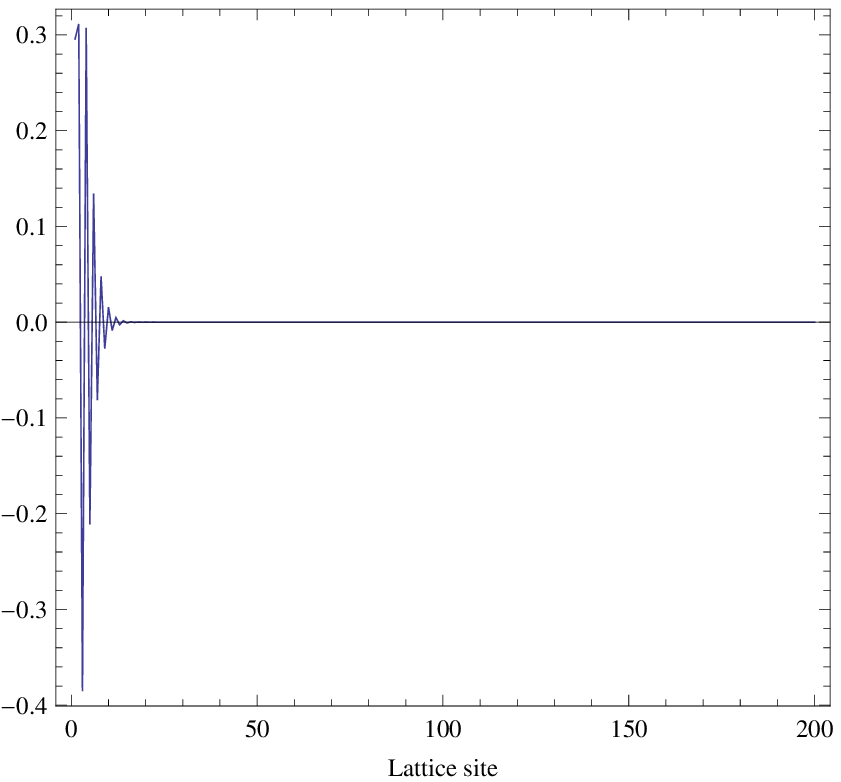}
}
\subfigure[] { \label{fig:d}
\includegraphics[width=0.45\columnwidth]{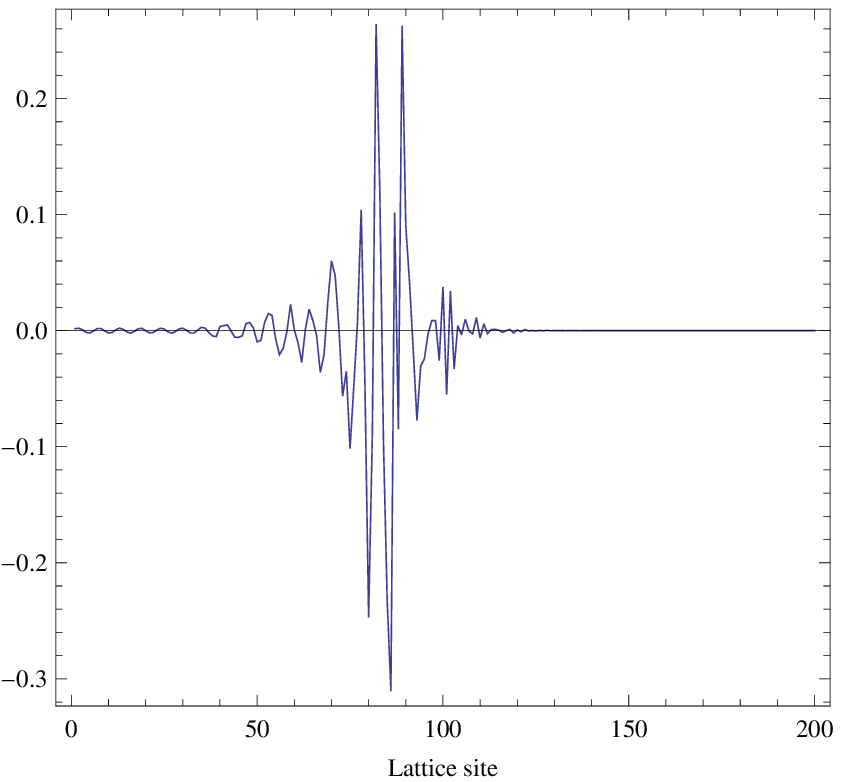}
}
\caption{(a) The energy spectrum of $H(k_x, k_y, k_z=0)$, with $p=q=1, h=-3/2$. (b) The energy spectrum of $H(k_x, k_y, k_z=\pi)$, with $p=q=1, h=-3/2$. (c) The edge mode of $H(k_x, k_y, k_z=0)$, with $p=2,q=1, h=-3/2$. (d) The bulk mode of $H(k_x, k_y, k_z=0)$, with $p=2,q=1, h=-3/2$. }
\label{H1}
\end{figure}

\section{Numerical results of edge states}
\label{num}

It is well known that the topological property usually reflected in two ways. One is the topological index defined in the bulk system, the other is the topological protected edge state on the boundary. This is the so called bulk-boundary correspondence. We will numerically demonstrate the existence of the edge mode on the boundary of nontrivial dimensional reduced Hopf insulator. To calculate the edge states, we have to solve the eigenvalues of the dimensional reduced Hopf Hamiltonian on a cylinder instead of 2D torus. That is, the system has periodic boundary condition in the y-direction and open boundary in the x-direction. Since it's still periodic in the y-direction, we can define the partial Fourier transformation
\be
c_{k_y}=\frac{1}{\sqrt{L_y}}\sum_y c(x,y) e^{ik_yy}
\ee
where $(x,y)$ denotes the coordinates of square lattice sites.

\begin{figure} \centering
\subfigure[] { \label{H2:a}
\includegraphics[width=0.45\columnwidth]{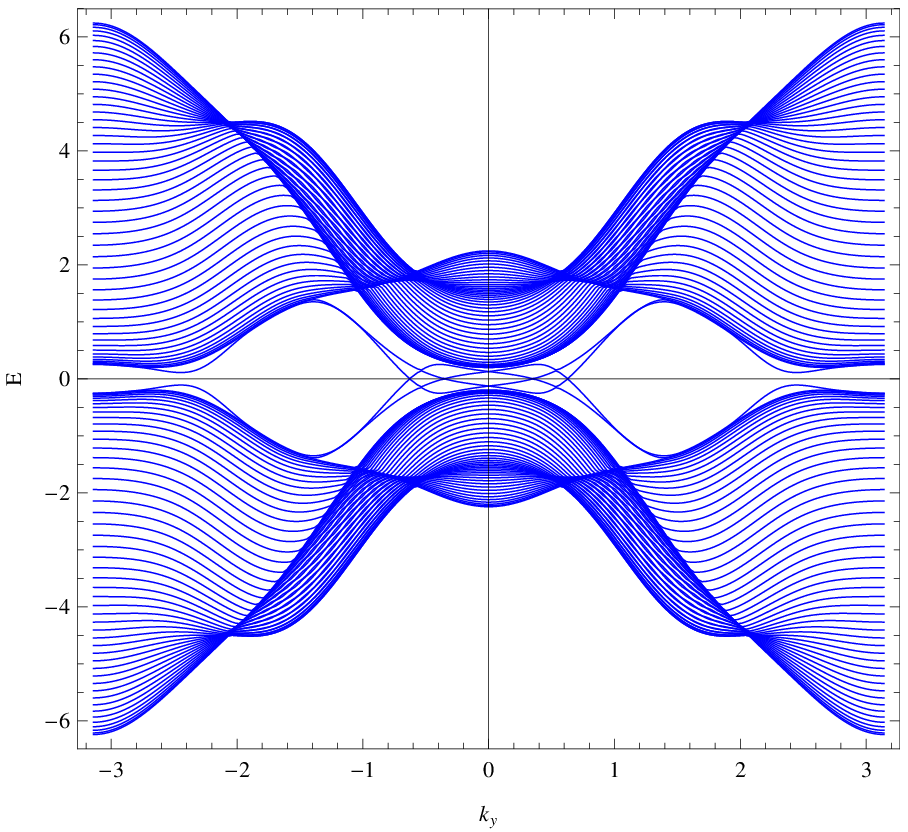}
}
\subfigure[] { \label{H2:b}
\includegraphics[width=0.45\columnwidth]{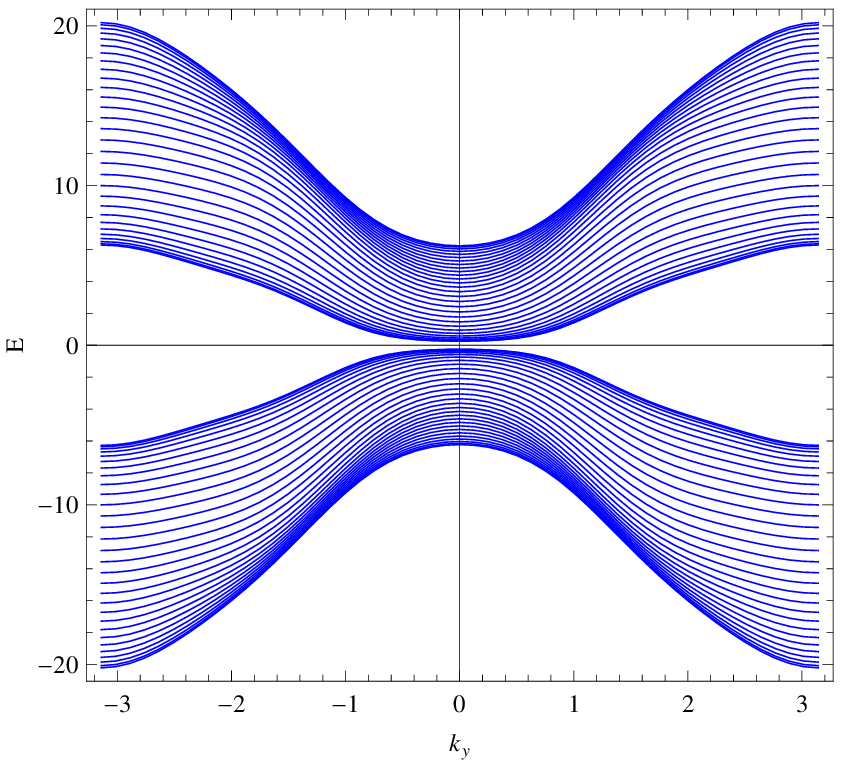}
}
\subfigure[] { \label{H2:c}
\includegraphics[width=0.45\columnwidth]{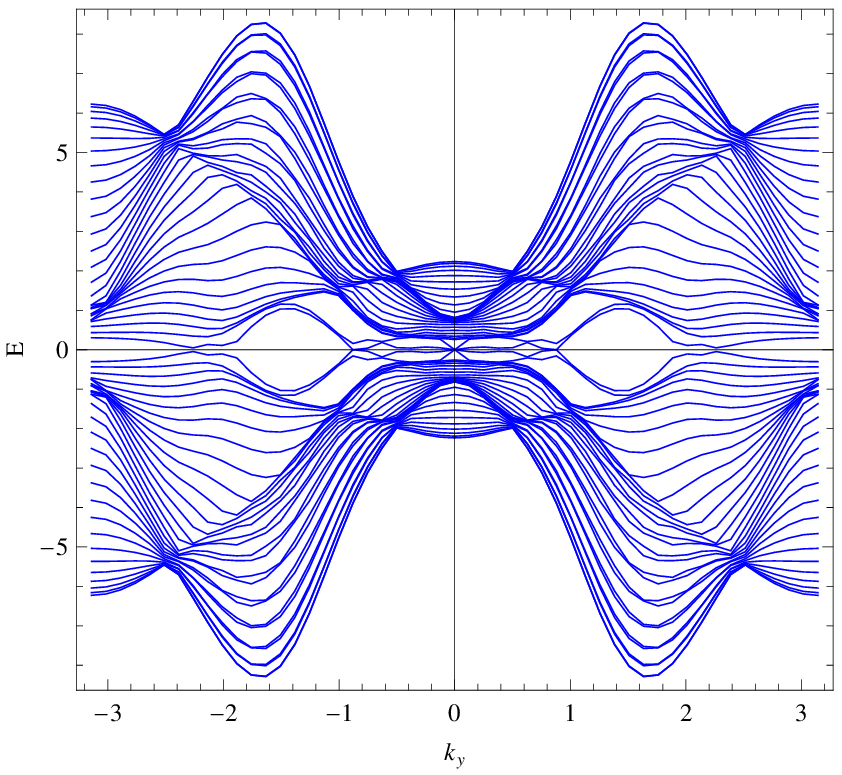}
}
\subfigure[] { \label{H2:d}
\includegraphics[width=0.45\columnwidth]{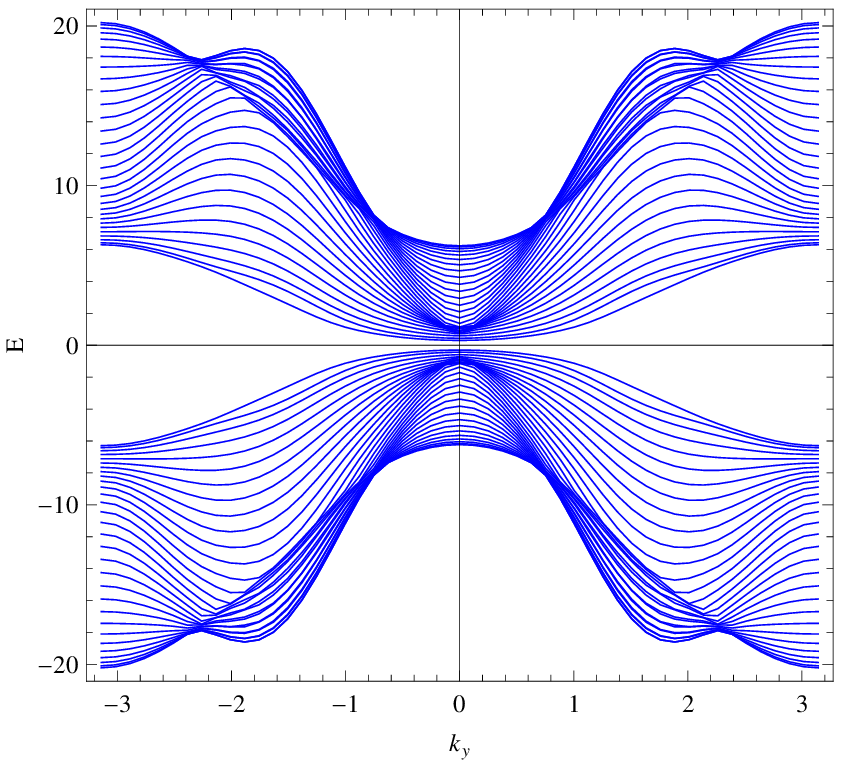}
}
\caption{(a) The energy spectrum of $H(k_x, k_y, k_z=0)$, with $p=2,q=1, h=-3/2$. (b) The energy spectrum of $H(k_x, k_y, k_z=\pi)$, with $p=2,q=1, h=-3/2$. (c) The energy spectrum of $H(k_x, k_y, k_z=0)$, with $p=3,q=1, h=-3/2$. (d) The energy spectrum of $H(k_x, k_y, k_z=\pi)$, with $p=3,q=1, h=-3/2$.}
\label{H2}
\end{figure}

We consider the cases with $\mathcal{H}=1$, which corresponds to the parameters $p=1, q=1, h=-3/2$. The dimensional reduced Hopf Hamiltonian can be written as
\begin{align}
&H=\sum_{k_y, x}\Big[t_1c^{\dagger}_{k_y,x}c_{k_y,x+1}+t_2c^{\dagger}_{k_y,x}c_{k_y,x+2}+h.c.\Big]\nonumber\\
&\qquad\qquad+t_0c^{\dagger}_{k_y,x}c_{k_y,x}\equiv\sum_{k_y}H_{1D}(k_y)\\
&t_0=2\sin k_y(\cos k_y+m)\sigma^x+[\sin^2k_y-(\cos k_y+m)^2]\sigma^z\nonumber\\
&t_1=\sin k_y\sigma^x+i(\cos k_y+m)\sigma^y-(\cos k_y+m)\sigma^z\nonumber\\
&t_2=\frac{i\sigma^y-\sigma^z}{2}\nonumber
\end{align}
Thus the 2D Hamiltonian is reduced to the sum of $L_y$ 1D tight-binding chains, here $L_y$ is the number of sites along $y$ direction. We can numerically solve the eigenvalues of the $H_{1D}(k_y)$ for each $k_y$. The result is shown in Fig. \ref{H1}. There are two curves crossing in $k_y=0$, which is the edge mode we seeking. Similarly, We can solve the eigenvalues of the case of $p=1, q=1, h=-3/2, k_z=\pi$ with the result shown in Fig. \ref{fig:b}. In this case, there are no edge modes, thus this state is topologically equivalent to vacuum state as we already claimed in section \ref{reduce}. This is exactly consistent with the conclusion we find with dimensional reduction. For the odd Hopf index, the two dimensional reduced models are in different classes, one nontrivial and the other trivial. In Fig.(\ref{fig:c}) and Fig. (\ref{fig:d}), we show the eigenstates of typical edge modes and bulk modes.

The band structure becomes more complicated for larger Hopf index. Using the same method, we solve eigenvalues for the $\mathcal{H}=2$ case and the results are shown in Fig. \ref{H2}. In Fig. \ref{H2:a}. there are four band crossing points. But they are symmetric about $k_y=0$. Therefore the left points will cancel the right points leaving no band crossing points. Hence, for the case with $\mathcal{H}=2$, $H(k_x, k_y, k_x=0)$ is topologically equivalent to $H(k_x, k_y, k_x=\pi)$, which has no edge states as shown in \ref{H2:b}.

The same argument can also be applied to the case with $\mathcal{H}=3$. In Fig. \ref{H2:c}, the left band crossing points will cancel the right band crossing points and there is only one band crossing point left. This suggests that the dimensional reduced Hopf insulator with $\mathcal{H}=3$ is topologically equivalent to $H(k_x, k_y, k_x=0)$, with $\mathcal{H}=1$. Also there are no edge states in Fig. \ref{H2:d}. This picture is exactly consistent with the discussions of dimensional reduction.

\section{conclusion}
\label{conclude}

In summary, we have found a $\mathbb{Z}_2$ classification for 2D two-band model with vanishing Chern number by dimensional reducing the Hopf insulator. This type of topological state is characterized by a $\mathbb{Z}_2$ index. Although it looks quite similar to the $\mathbb{Z}_2$ index of quantum spin Hall state, the dimensional reduced Hopf insulator is not protected by any symmetry. Therefore it is an exception to the periodic table of topological insulator just like Hopf insulator. We also construct explicit tight-banding Hamiltonian with nontrivial $\mathbb{Z}_2$ index and demonstrate the existence of the edge modes. It is possible that the tight-banding model we have proposed can be experimentally realized by certain orbital or spin dependent hopping in certain material.

\end{document}